\documentclass[
 reprint,
 amsmath,amssymb,
 aps,
]{revtex4-2}

\usepackage{graphicx}
\usepackage{dcolumn}
\usepackage{bm}
\usepackage{comment}
\usepackage{textcomp}
\usepackage{ulem}
\usepackage[]{graphicx,xcolor}
\usepackage{amsmath}
\usepackage{bbold}
\usepackage[utf8]{inputenc}
\usepackage[T1]{fontenc}
\usepackage[english]{babel}
\usepackage{verbatim}
\usepackage{textcomp}
\usepackage{ulem}
\usepackage[]{graphicx,xcolor}
\usepackage{soul}

\newcommand{\p}{\partial}

\DeclareMathOperator{\sign}{sign}

\DeclareMathOperator{\diag}{diag}

\begin{document}

\preprint{APS/123-QED}

\title{Dyakonov surface waves in dielectric crystals with negative anisotropy}
\author{Dmitry~A.~Chermoshentsev}
\email[]{e-mail: dmitry.chermoshentsev@skoltech.ru}
\affiliation{Skolkovo Institute of Science and Technology, 143025 Moscow Region, Russia}
\affiliation{Moscow Institute of Physics and Technology, 141701 Institutskiy pereulok 9, Moscow Region, Russia}
\affiliation{Russian Quantum Center, Skolkovo, Moscow 143025, Russia}
\author{Evgeny~V.~Anikin}
\affiliation{Skolkovo Institute of Science and Technology, 143025 Moscow Region, Russia}
\author{Sergey~A.~Dyakov}
\email[]{e-mail: s.dyakov@skoltech.ru}
\affiliation{Skolkovo Institute of Science and Technology, 143025 Moscow Region, Russia}
\author{Nikolay~A.~Gippius}
\affiliation{Skolkovo Institute of Science and Technology, 143025 Moscow Region, Russia}
\date{\today}

\begin{abstract}

We report the prediction of a new type of Dyakonov surface waves that propagate along the flat strip of the interface between two dielectrics with negative anisotropy. It is shown that the surface waves condition is satisfied for negatively anisotropic dielectrics due to specific boundaries of the strip waveguide confined between two metallic plates. Such modes are studied by the perturbation theory in the approximation of weak anisotropy. The existence of Dyakonov surface waves in negative uniaxial crystals motivates us to reconsider the list of materials suitable for their practical implementation. We believe that this work opens a new unexplored research area in the field of surface waves.
\end{abstract}
\maketitle

\section{Introduction}

Dyakonov surface waves (DSWs) is an important class of electromagnetic surface waves which exist at the interface of two dissimilar materials, at least one of which is anisotropic. In contrast to surface plasmon polaritons, DSWs have no theoretical limit in propagation length as they can exist at the interface of two lossless dielectrics. As shown by M. Dyakonov and N.\,S.\, Averkiev in late 80s \cite{Dyakonov1988, averkiev1990electromagnetic}, this type of DSWs can propagate in a certain range of angles relative to the optical axis and only in a system with positively anisotropic material(s). A positive optical anisotropy means that $\varepsilon_{\parallel}>\varepsilon_\perp$, where $\varepsilon_{\parallel}$ (or $\varepsilon_\perp$) is the principal value of dielectric permittivity tensor parallel to (or perpendicular to) the optical axis. Since then, extensive research has been performed toward the theoretical studies of Dyakonov-like surface waves at interfaces of different combinations of isotropic, uniaxial, biaxial, and chiral materials with positive anisotropy \cite{Narimanov2018,Lakhtakia2007,Gao:09,repan2020wave, zhang2020unusual, karpov2019dyakonov,fedorin2019dyakonov, Mackay2019, Lakhtakia2020, Lakhtakia2020a,zhou2020theory, Ardakani2016,Moradi2018,Takayama2018,  Alshits2002,  Nelatury:07, Darinskii2001, Furs2005, Furs2005a,PoloJr.2007}. A narrow range of propagation angles and the requirement of positive anisotropy sufficiently decreases the number of materials suitable for practical realization of DSWs. Nevertheless, experimental observation of such waves was demonstrated by O.\,Takayama et al. for the interface formed by a dielectric liquid, and biaxial KTP crystal \cite{Takayama2009}. Later, the same group have experimentally demonstrated the existence of Dyakonov-like guided modes in thin aluminum oxide nanosheets placed between anisotropic crystal and dielectric liquid \cite{Takayama2014}.
 
\begin{figure}[b!]
    \centering
    \includegraphics[width=0.9\linewidth]{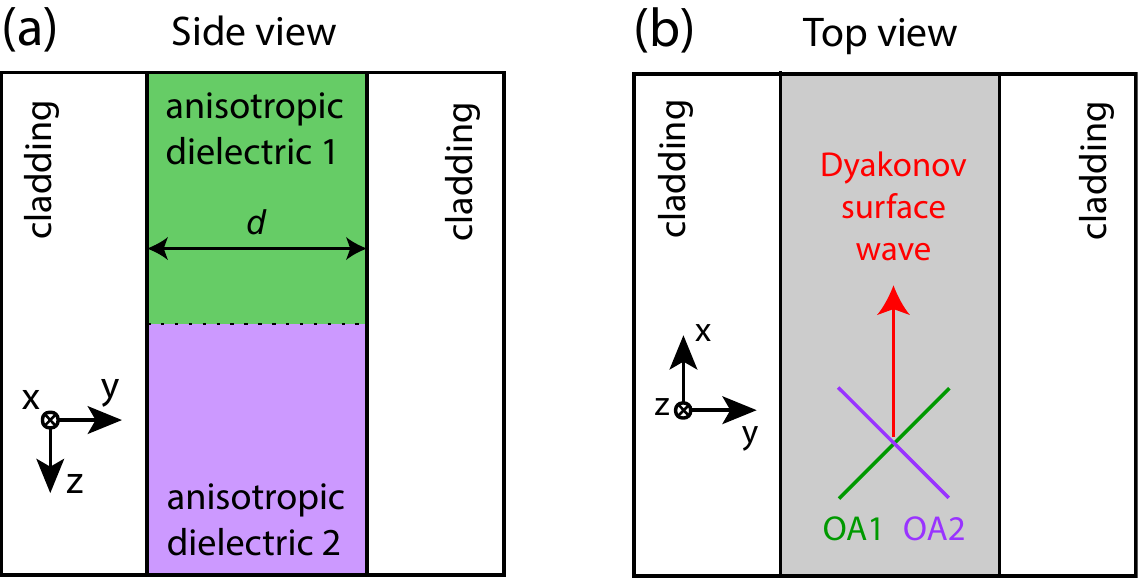}
    \caption{(a) Side view and (b) top view of the Dyakonov waveguide. Optical axes (OA) of anisotropic dielectrics 1 and 2 are perpendicular to each other and form the angle of $45^\circ$ to the waveguide boundaries.}
    \label{fig:geometry}
\end{figure}

The existence of DSWs has been also theoretically predicted for interfaces between isotropic material and metamaterial with artificially designed anisotropy \cite{Chiadini2016, Lakhtakia2007, Lakhtakia2014, Pulsifer2013, artigas2005dyakonov, jacob2008optical, Takayama2012, Takayama2012coupling, Cheng2021}. In such structures the angular existence domain of DSWs can be sufficiently extended. The experimental realization of such DSWs has been reported in Ref.\,\cite{takayama2017midinfrared}.

Very recently, a practically important case of Dyakonov-like surface waves in finite-size resonator structures has been studied for cylindrical waveguides \cite{Kajorndejnukul2019, Golenitskii2019} as well as for flat interfacial strip waveguides \cite{Chermoshentsev2020, Anikin2020}. It has been demonstrated that due to a non-zero curvature of cylindrical waveguides, Dyakonov surface waveguide modes (DSWMs) in them inevitably have radiative losses. At the same time, in flat interfacial strip waveguides such modes can propagate without radiative losses like classical DSWs. 



Because in the pioneering works \cite{Dyakonov1988, averkiev1990electromagnetic} it has been shown that the main requirement for materials to support DSWs at their infinite interfaces is being positevely anisotropic, DSWs community have been focused on positive crystals and up to now there are no literature data on DSWs in negative anisotropic crystals. In this work we demonstrate that the interface between two negative anisotropic crystals can support DSWs, if it is confined between two metallic plates.

\section{Low anisotropy approximation}

\begin{figure*}[th]
    \centering
    \includegraphics[width=1\linewidth]{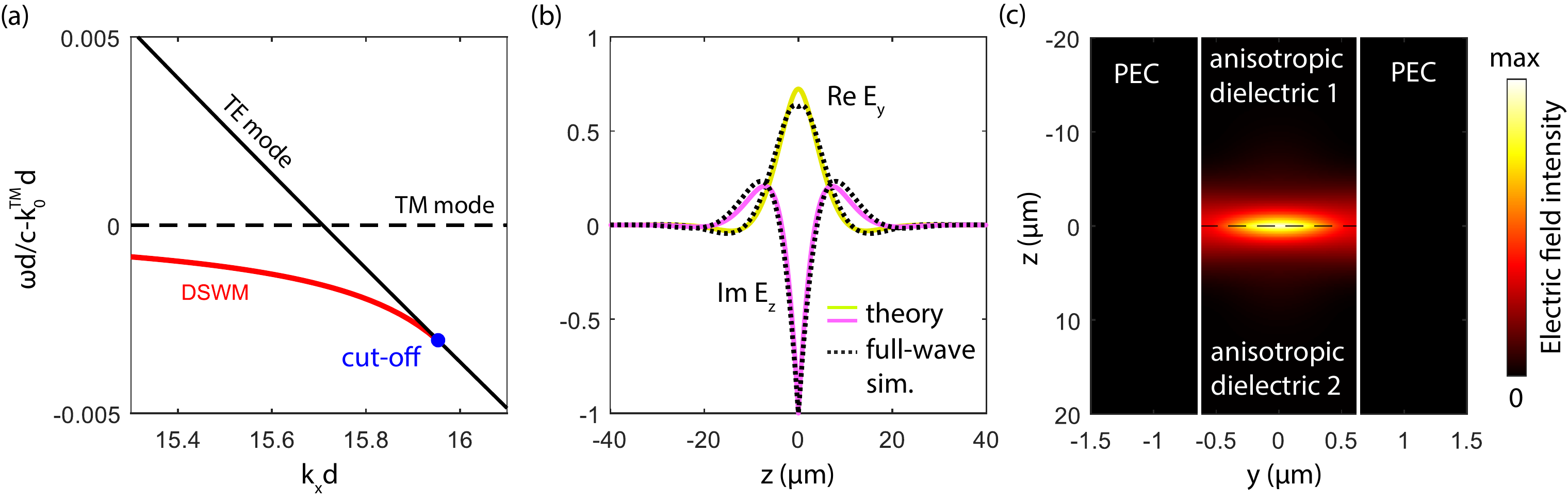}
    \caption{(a) The dispersions of the DWM (red line) and the TE and TM waveguide modes of an anisotropic waveguide (black solid and dashed lines) shown as a difference with the dispersion of the TM waveguide mode. (b) Theoretically calculated fields $E_y(0, z)$ and $E_z(0, z)$ (yellow and magenta lines) shown together with COMSOL simulation results (black dashed lines). (c) Theoretically calculated electric field intensity in DWM. Please note that in panel (c) the limits of $y$- and $z$-axes are different. Calculations in (b) and (c) are made for $\lambda = 1550$\,nm, $k_xd =  15.705$ and $d = 1265$\,nm. Dielectric permittivities for all panels are $\varepsilon_o =9.75$, $\varepsilon_e = 9.0$, $\delta\varepsilon=0.375$.} 
    \label{fig:disp}
\end{figure*}

The structure under consideration consists of two spliced dielectric slabs confined by thick metallic plates at $|y|=d/2$ (Fig.\ref{fig:geometry}a,b). The slabs are infinite in $x$-direction and semi-infinite in $z$-direction; the interface between them is at $z=0$. The principal values of the permittivity tensor of both slabs are $\varepsilon_{e}, \varepsilon_{o}, \varepsilon_{o}$ with $\varepsilon_{e}<\varepsilon_{o}$, i.e. we consider the case of a negative anisotropy. The values of $\varepsilon_{e}, \varepsilon_{o}$ are assumed to be real and positive. The optical axes of the upper and lower slabs are rotated by an angle of $\pm45^{\circ}$ about the $z$-axis as shown in Fig.\ref{fig:geometry}b.


We represent the dielectric tensor $\hat{\varepsilon}(z)$ inside the waveguide as $\hat{\varepsilon}(z) = \hat{\varepsilon}_0 + \delta\hat{\varepsilon}(z)$, where
\begin{equation}
\label{initialapp}
    \begin{gathered}
    \hat{\varepsilon}_0 = \diag(\varepsilon_1, \varepsilon_1, \varepsilon_2), \,\,\,\,\,\,\,
    \delta\hat{\varepsilon}(z) = \sign(z)
                    \begin{pmatrix}
                        0 & \delta\varepsilon & 0\\
                        \delta\varepsilon & 0 & 0\\
                        0 & 0 & 0
                    \end{pmatrix}
    \end{gathered}
\end{equation}
with $\varepsilon_{1}=\frac{\varepsilon_{o}+\varepsilon_{e}}{2},\ \varepsilon_{2}=\varepsilon_{o}$ and $\delta\varepsilon=\varepsilon_{2}-\varepsilon_{1}$. In addition to this model we use two important approximations which allows us to develop a theory for DSWMs in negative anisotropic crystals. First, we assume that materials have low anisotropy, $\delta\varepsilon\ll\varepsilon_{1,2}$ (the most often case for natural anisotropic crystals), that allows us to consider the Dyakonov waveguide as a small perturbation of a homogeneous anisotropic waveguide with a dielectric tensor $\hat\varepsilon_{0}$. Second, we replace metallic plates with a perfect electric conductor (PEC) plates, which means that the metallic cladding can be accounted for simply by using specific PEC boundary conditions. For the sake of generality, in the end of this paper, we will show that DSWMs exist in a more general case when these two approximations are not valid.




The waveguide modes of unperturbed anisotropic waveguide with the PEC walls can be analytically described for arbitrary $k_{x}$ and $k_{z}$ as shown in Supplemental Materials. 
The dispersion curves of the lowest TM- and TE-modes at $k_z=0$ are set by the following equations:
\begin{equation}\label{coefficients1}
    \begin{gathered}
        (\omega^{\scriptscriptstyle \mathrm{TE}}_{k_x,0})^2=\frac{c^2}{\varepsilon_{2}}\left[\left(\frac{\pi}{d}\right)^2+k_{x}^2\right],\,\,\,\,\,\,\,\,
        (\omega^{\scriptscriptstyle \mathrm{TM}}_{k_x,0})^2= \frac{c^2 k_x^2}{\varepsilon_{1}}
    \end{gathered}.
\end{equation}
One can see from Eqs.\,\ref{coefficients1} that these modes intersect and the wavevector of intersection is the following: 
\begin{equation}\label{intersection}
    k_{x}=\frac{\pi}{d}\sqrt{\frac{\varepsilon_{1}}{\varepsilon_{2}-\varepsilon_{1}}}.
\end{equation}
It is important to note that the intersection of the lowest TE- and TM-modes of the unperturbed uniaxial waveguide with $\varepsilon_1<\varepsilon_2$ occurs only due to specific PEC boundary conditions. We will show
that close to the intersection, the perturbation $\delta\hat{\varepsilon}$ leads to considerable mixing of TE- and TM-modes giving rise to a special type of the surface mode. Hereinafter, we call this electromagnetic mode as Dyakonov surface wave of the II type (DSW-II), while a classical Dyakonov surface wave is referred to as Dyakonov surface wave of the I type (DSW-I). DSW-II does not exist at the infinite interface of two negatively anisotropic plates; it also does not exist at the strip interface of two spliced negatively anisotropic slabs confined between air plates. 

\begin{figure}[b!]
    \centering
    \includegraphics[width=1\linewidth]{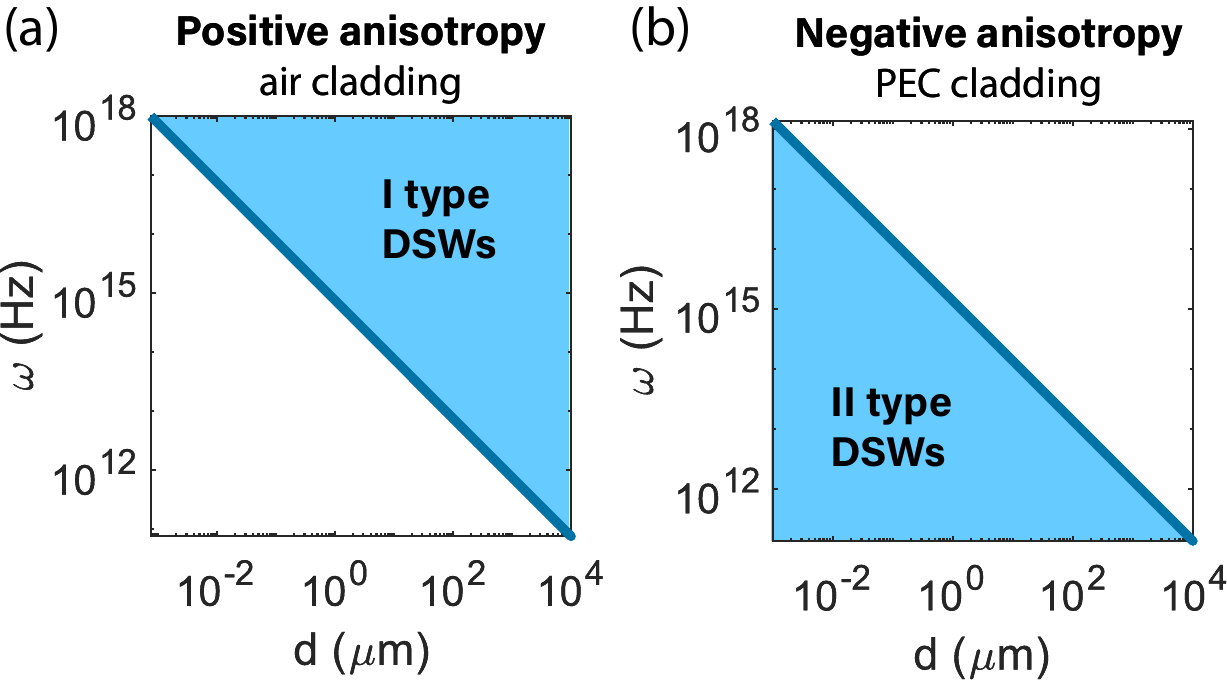}
    \caption{Domains of existence of (a) the I type and (b) the II type DSWs in the structure shown in Fig.\,\ref{fig:geometry} with dielectrics with positive or negative anisotropy. In (a) $\varepsilon_o=9.0$, $\varepsilon_e=9.75$, in (b) $\varepsilon_o=9.75$, $\varepsilon_e=9.0$.}
    \label{fig:domains}
\end{figure}

In the approximation of a low anisotropy, the intersection of the dispersion curves of the lowest TM and TE modes is far from of the next order waveguide modes (see Supplemental Materials). It enables us to expand the field of DSW-II over the lowest TE and TM modes of unperturbed waveguide: 
\begin{equation}
    \label{field_expansion}
    \vec{E}(y,z) =\int \frac{dk_z}{2\pi} \left[ \alpha(k_z) \vec{E}^{\mathrm{\scriptscriptstyle TE}}_{k_x,k_z}(y) + 
                \beta(k_z) \vec{E}^{\mathrm{\scriptscriptstyle TM}}_{k_x,k_z}(y) \right]e^{ik_{z}z}
\end{equation}
where $\alpha(z)$ and $\beta(z)$ are the slowly varying envelopes and $\vec{{E}}^{\mathrm{\scriptscriptstyle TE}}_{k_x,k_z}$ and $\vec{{E}}^{\mathrm{\scriptscriptstyle TM}}_{k_x,k_z}$ are the normalized lowest TE and TM modes, which are described as follows:
\begin{equation}
    \begin{gathered}
    \vec{E}^{\mathrm{\scriptscriptstyle TE}}_{k_x,k_z}(y) = \sqrt{\frac{2}{d}}\frac{1}{\varepsilon_{2}^{3/2}\varepsilon_{1}}\frac {c^2}{\omega^2}\begin{pmatrix}
                   \varepsilon_{2}k_{x}k_{z}\cos{\frac{\pi}{d}y}\\
                    i\varepsilon_{2}\frac{\pi}{d} k_{z}\sin{\frac{\pi}{d} y}\\
                    -\varepsilon_{1}[k_{x}^2+(\frac{\pi}{d})^2]\cos{\frac{\pi}{d} y}
              \end{pmatrix}\\
    \vec{E}^{\mathrm{\scriptscriptstyle TM}}_{k_x,k_z}(y) = \frac{1}{\varepsilon_{1}\sqrt{d}}\frac {c}{\omega}\begin{pmatrix}
                   0\\
                    k_{x}\\
                    0
              \end{pmatrix}
    \end{gathered}
\label{TETM_field}              
\end{equation}
Below we present the result of applying the perturbation theory for finding the expressions of the amplitudes $\alpha(z)$ and $\beta(z)$ and the dispersion for DSW-II (see Supplemental Materials for details on perturbation theory). After plenty of algebraic manipulations one can obtain

\begin{equation}
    \label{exact_envelopes}
    \begin{gathered}
        \alpha(z) =\frac{\omega^2\sigma\delta\varepsilon e^{-\kappa_2|z|}}
            {\left( \gamma^{\mathrm{\scriptscriptstyle TE}}_{k_{x},0} - \frac{\kappa_2^2}{2m_{1}}\right) }
                     -   \frac{\omega^2\sigma\delta\varepsilon e^{-\kappa_1|z|}}
            {\left(\gamma^{\mathrm{\scriptscriptstyle TE}}_{k_{x},0} - \frac{\kappa_1^2}{2m_{1}}\right)}\\             
        \beta(z) = -i\left(\frac{e^{-\kappa_2|z|}}{\kappa_2} - 
            \frac{e^{-\kappa_1|z|}}{\kappa_1}
            \right)
    \end{gathered}, 
\end{equation}
where decay constants $\kappa_{1,2}$ have positive real and are found as two roots of the characteristic equation
\begin{equation}
    \label{char_eq}
    \left(\gamma^{\mathrm{\scriptscriptstyle TE}}_{k_{x},0} - \frac{\kappa^2}{2m_{1}}\right) \left(\gamma^{\mathrm{\scriptscriptstyle TM}}_{k_{x},0} - \frac{\kappa^2}{2m_{2}}\right) + \omega^4\sigma^2\delta\varepsilon^2\kappa^2 = 0,
\end{equation}
where 
\begin{equation}\label{coefficients}
    \begin{gathered}
        \sigma=\frac{2\sqrt{2 \varepsilon_{2}}}{\varepsilon_{1}^{3/2} \pi}\frac{k_{x}}{k_{x}^2+(\frac{\pi}{d})^2},\,\,\,\,\,\,\,\,
        m_{1}^{-1}=m_{2}^{-1}=\frac{2c^2}{\varepsilon_{1}}
    \end{gathered}.
\end{equation}
In Eqn.\eqref{char_eq} we use the notation $\gamma_{k_x,k_z}^{\mathrm{\scriptscriptstyle TE(TM)}}=\left(\omega_{k_xk_z}^{\mathrm{\scriptscriptstyle TE(TM)}}\right)^2 - \omega^2$ for brevity.


The dispersion equation of the surface mode can be obtained from the boundary conditions and have the following form:
\begin{equation}
\label{dm_dispersion}
    \sqrt{m_1m_2\gamma^{\mathrm{\scriptscriptstyle TE}}_{k_{x},0}\gamma^{\mathrm{\scriptscriptstyle TM}}_{k_{x},0}}
    =   2m_1m_2(\omega^2\sigma\delta\varepsilon)^2 -m_2\gamma^{\mathrm{\scriptscriptstyle TM}}_{k_{x},0}
\end{equation}
It turns out that the transcendental equation (\ref{dm_dispersion}) has a solution (red line in Fig.\,\ref{fig:disp}a) and, hence, the DSW-II exists. As expected, the DSW-II dispersion curve is close the intersection of the lowest TE- and TM-modes. One can see from Fig.\,\ref{fig:disp}a that there is an upper cut-off for the DSW-II which is due to the fact that the 
solution of Eq.~\eqref{dm_dispersion} exists when the right-hand side is non-negative.
In the cut-off point, the DSW-II transforms into the TE-mode and becomes completely delocalized. The existence of the upper cut-off point of the dispersion curve leads to fact that there is an upper cut-off waveguide thickness $d$ at which the DSW-II cease to exist. After substitution of the coefficients (\ref{coefficients}) into the right hand side of Eq.~\eqref{dm_dispersion} we obtain that the cut-off thickness depends on the frequency $\omega$ or vacuum wavelength $\lambda$ as follows:
\begin{equation}\label{cut_off_d}
    d\le\frac{c\pi}{\omega}\sqrt{\frac{\varepsilon_{1}}{\varepsilon_{2}\delta\varepsilon}}=\frac{\lambda}{2}\sqrt{\frac{\varepsilon_{1}}{\varepsilon_{2}\delta\varepsilon}}.
\end{equation}
In Ref.\,\cite{Anikin2020}, where a DSW-I in positively anisotropic materials between two air plates are considered, the situation is opposite. Namely, there is a lower cut-off point for the dispersion curve. Fig.\,\ref{fig:domains} visually illustrates the domains of existence of DSW-I and DSW-II in terms of the waveguide thicknesses $d$ for different frequencies $\omega$. Since for DSW-II there is the upper cut-off waveguide thickness, DSW-II does not exist in the case of an infinite ($d\to\infty$) interface between two anisotropic materials, in contrast to DSW-I. 

The distribution of electric field intensity of DSW-II within the waveguide's vertical cross-section calculated by formula \eqref{field_expansion} is shown in Fig.\,\ref{fig:disp}c. One can see that the field decays away from the interface and has the maximum in the middle of the waveguide. To verify the developed theory, we calculate the $z$-dependence of electric field by formula \eqref{field_expansion} and compare it with the results of full-wave electromagnetic simulations made in COMSOL (Fig.\,\ref{fig:disp}b). We obtain that apart from a generally very good agreement between two field profiles, there is the slight discrepancy between two curves near the interface  which is caused by the influence of the higher-order waveguide modes of the unperturbed waveguide (see Supplemental Materials for details). 

\begin{figure*}[t!]
    \centering
    \includegraphics[width=1\linewidth]{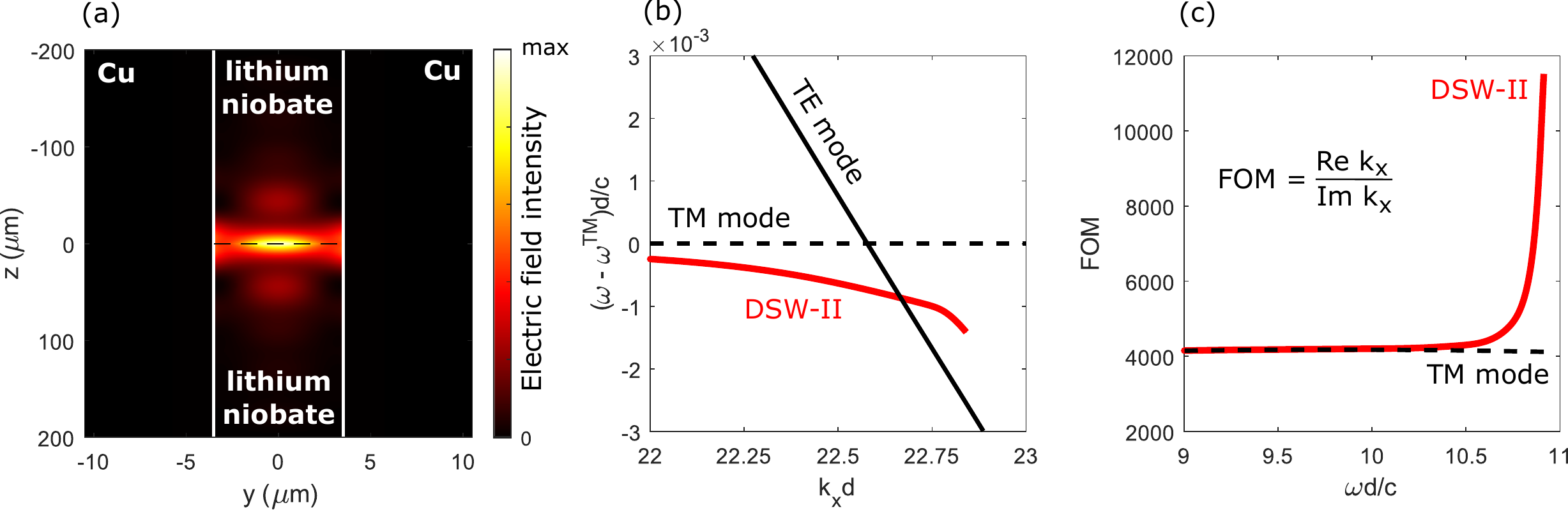}
    \caption{(a) Electric field intensity of DSW-II calculated in LiNbO$_3$/Cu structure at the vacuum wavelength $\lambda = 4.04$\,$\mu$m. Please note that in panel (a) the limits of $y$- and $z$-axes are different. (b) Dispersions of DSW-II, TM hybrid mode and TE guided mode. (c) Figure-of-merit for DSW-II and TM hybrid mode as a function of photon energy. Calculations for all panels are made at the waveguide thickness $d=7$\,$\mu$m. Optical axes of the upper and lower LiNbO$_3$ plates are oriented as shown in Fig.\,\ref{fig:geometry}b. Please note that in panels (b) and (c) the cut-off points for DSW-II are determined approximately due to limitation of the computational domain size in COMSOL. Dielectric parameters of LiNbO$_3$ and Cu are taken from \cite{babar2015optical, zelmon1997infrared}.} 
    \label{fig:real}
\end{figure*}

As was previously mentioned, the model described above is developed for the low anisotropy limit. Nevertheless, the analytical description of DSWs-II can also be made for the arbitrary anisotropy in the approximation of a thin  waveguide $d\ll\lambda$. In this limit, only the lowest TM mode exists with the electric field components $E_x$ and $E_z$ being small. As a result, a surface-wave-like solution of Maxwell's equations can be analytically obtained (see Supplemental materials for details) and the corresponding propagation constant is the following:
\begin{equation}\label{eq:dispersion_thin_wg}
    k_x = \sqrt{k_0^2\varepsilon_{1} + \kappa^2} \approx 
        k_0\sqrt{\varepsilon_{1}} + \frac{\kappa^2}{2k_0\sqrt{\varepsilon_{1}}}
\end{equation}
where $\kappa$ is described as follows:
\begin{equation}\label{eq:ey}
    \kappa = \frac{k_0^4 d^3\delta\varepsilon^2}{3\sqrt{3}}
           \left(\sqrt{\frac{\varepsilon_{2}}{\varepsilon_{1}}} - 1\right)
\end{equation}

\section{II-type DSW in real materials}

So far we have studied DSWs-II in the dispersionless anisotropic materials confined between two idealized PEC plates. Let us take one step towards practical implementation of DSW-II and consider a structure consisting of lithium-niobate-based Dyakonov waveguide placed between two copper plates. We employ lithium niobate (LiNbO$_3$) because it is lossless negative anisotropic material that is widely used in various linear and non-linear optical applications including optical waveguides. The COMSOL simulation results for the modes of this structure are shown in Fig.\,\ref{fig:real}. First, we note that DSW-II in the LiNbO$_3$/Cu structure exists, and like in the ideal case, the cross-sectional field profile (Fig.\,\ref{fig:real}c) has a global maximum at the interface between upper and lower LiNbO$_3$ waveguides. There are also local maxima of the field distribution at some distance from the interface. Further, in contrast to the ideal case, the initial TM mode of LiNbO$_3$/Cu structure is a hybrid mode resulting from the mixing between the surface plasmon polaritons at LiNbO$_3$/Cu interface and the TM guided mode of LiNbO$_3$ waveguide. Due to more complex mode structure of the LiNbO$_3$/Cu waveguide in comparison to the ideal case, the cut-off point of DSW-II does not lie on the TE guided mode dispersion curve anymore (Fig.\,\ref{fig:real}b). The hybrid TM mode evanescently decays into the absorptive Cu plate, thus having a non-zero imaginary part of the propagation constant. Eventually, this leads to the fact that in the considered example, the DSW-II has absorption losses. The corresponding propagation length of DSW-II can be quantitatively estimated from the figure-of-merit calculated as:
\begin{equation}
    \mathrm{FOM} = \frac{\mathrm{Re}  \:k_x}{\mathrm{Im} \:k_x}
\end{equation}
which has a meaning of the decay length expressed in the units of the DSW-II wavelength. As one can see from Fig.\,\ref{fig:real}b, the propagation length of the DSW-II exceeds that of the hybrid TM mode and increases when approaching the cut-off point. In the far-infrared range or radiowaves, the skin depth in metals become very small and much larger propagation length of DSW-II can be expected. Another approach for increasing the propagation length of DSW-II can be based on using photonic crystals as cladding. 

\section{Conclusions}
In conclusion, we have studied a new type of Dyakonov surface waves that propagate along the flat strip of the interface between two uniaxial dielectrics with negative anisotropy. We have shown that the surface waves condition is satisfied for negatively anisotropic dielectrics due to specific boundaries of the strip interfacial waveguide confined between two metallic plates. We have studied such modes using the perturbative approach under the assumption of weak anisotropy. We generalized our results made for idealized materials to the case of real materials and found out that DSWs-II should have a finite propagation length due to absorption in metal. We suggest that the propagation length of DSWs-II can be increased by switching to FIR or radiowaves or by using photonic crystals as cladding. Finally, the existence of Dyakonov surface waves in negative uniaxial crystals motivates us to reconsider the list of materials suitable for the practical realization of Dyakonov surface waves. Thus, we believe that this work opens a new unexplored research area in the field of surface waves.
\begin{acknowledgments}
This work was supported by the Russian
Foundation for Basic Research (Grant No. 18-29-20032).
\end{acknowledgments}

%

\end{document}